\documentclass[a4paper,11pt]{article}
\usepackage{jinstpub} % for details on the use of the package, please see the JINST-author-manual
\usepackage{lineno}
\usepackage{comment}
\title{\boldmath Scintillation of Ar/CF$_4$ mixtures: glass-THGEM characterization with 1\% CF$_4$ at 1-1.5~bar}

\author[a]{P. Amedo,}
\author[a,b]{R. Hafeji,}
\author[c]{A. Roberts,}
\author[c]{A. Lowe,}
\author[c]{S. Ravinthiran,}
\author[a]{S. Leardini,}
\author[c]{K. Majumdar,}
\author[c]{K. Mavrokoridis,}
\author[a]{D. González-Díaz}

\affiliation[a]{Instituto Gallego de Física de Altas Energías, Univ. de Santiago de Compostela, Campus sur, Rúa Xosé María Suárez Núñez, s/n, Santiago de Compostela, E-15782, Spain}
\affiliation[b]{Instituto de F\' isica Corpuscular (IFIC), CSIC \& Universitat de Val\`encia,\\ Calle Catedr\'atico Jos\'e Beltr\'an, 2, 46980 Paterna, Valencia, Spain}
\affiliation[c]{Department of Physics, University of Liverpool, Oliver Lodge Bld, Oxford Street, Liverpool L69 7ZE, UK}

\emailAdd{pablo.amedo.martinez@usc.es}

\abstract{Argon gas doped with 1\% wavelength-shifter (CF$_4$) has been employed in an optical time projection chamber (OTPC) to image cosmic radiation. We present results obtained during the system commissioning, performed with two stacked glass-THGEMs and an EMCCD camera at 1~bar. Preliminary estimates indicate that the combined optical gain was of the order of 10$^6$ (ph/e), producing sharp and high-contrast raw images without resorting to any filtering or post-processing. A first assessment of the impact of pressurization showed no change in the attainable gains when operating at 1.5~bar. }


\proceeding{LIDINE 2023: Conference in LIght Detection In Noble Elements\\
  20-22 September 2023\\
  Madrid, Spain}

\arxivnumber{2312.07503} % Only if you have one

\begin{document}
\maketitle
\flushbottom

\section{Introduction}
\label{sec:Introduction}

Operation of gaseous time projection chambers (TPCs) based on noble gases typically require the addition of some molecular additive. These are used to reduce the spatial spread and collection time of the primary ionization, to minimize photon and ion feedback and to attain greater stability. Some of the most well-known molecules are CH$_4$, CO$_2$ and CF$_4$, although more complex hydrocarbons are also employed. CF$_4$ displays some particularly interesting properties, such as low electron diffusion and scintillation in the 500-700 nm range \cite{Morozov2011,Morozov2010}. Argon/CF$_4$ mixtures have been shown to produce significant primary \cite{Amedo2023} and secondary \cite{Fraga2003} scintillation in the CF$_4$ band centered at around 630~nm, with as little as 1\% doping. This opens up the possibility of using these mixtures in connection with scientific CMOS/CCD cameras, that have good quantum efficiency in the visible range.

Weakly-doped argon mixtures may also prove useful in the context of neutrino physics, aiming for instance at the Deep Underground Neutrino Experiment (DUNE). The main target and detection medium for both the near and far detector complexes of DUNE is argon \cite{DuneVolI}. In particular, a high pressure argon-based TPC is foreseen to be located in the near detector complex, with the aim of reconstructing the low-energy hadron content of neutrino interactions down to a few MeV \cite{ND_CDR,NDGAR_Snowmass2021}. In this scenario, a mixture capable of providing both time tagging through primary scintillation and track imaging through secondary scintillation may significantly expand the capabilities of classical charge-based TPC readouts. Keeping any gas additive to a minimum, so as to keep the fraction of non-argon reactions to \%-level or below is essential. %Coupled with adequate amplification structures and, given the success of Timepix-based cameras\cite{ARIADNE_plus}, it could enable a full optical readout of the detector.

With this in mind, a systematic study of the optical gain of a double glass-THGEM stack \cite{glass_thgems} was performed in argon doped with 1\% CF$_4$  at around atmospheric pressure. The characterization was carried out using an alpha source and, when good amplification and transfer fields had been determined, imaging of cosmic radiation was performed.

\section{Experimental setup and procedure}
\label{sec:Setup}

\begin{figure}[h!!!]
\centering % \begin{center}/\end{center} takes some additional vertical space
\includegraphics[width=0.7\textwidth,origin=c,angle=0]{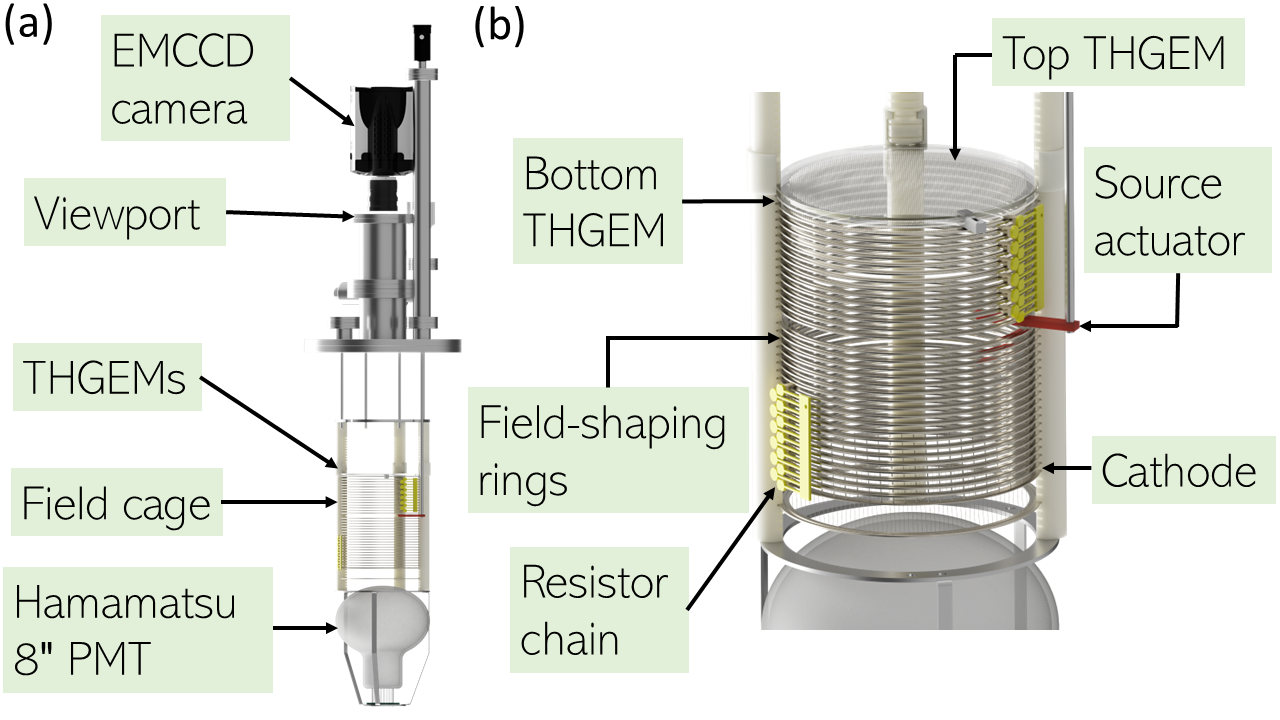}
% "\includegraphics" from the "graphicx" permits to crop (trim+clip)
% and rotate (angle) and image (and much more)
\caption{\label{fig:CAD} A 3D CAD design showing (a) the main components of the detector and (b) a closer view of the field cage region.}
\label{CAD}
\end{figure}

The experimental setup consisted of a cylindrical field cage housed within a 40 litre ultra-high vacuum (UHV) vessel (Fig.\ref{fig:CAD}). An Andor iXon Ultra 897 EMCCD camera with a Spacecom VF50095M lens was mounted on top of the centrally positioned DN 100 CF upper flange of the vessel. This top port was surrounded symmetrically by four ports with DN 40 CF flanges, which allow the mounting of electric and gas feedthroughs. The pressure and vacuum levels were monitored through one of these with a MKS 974B-71024 QuadMag Vacuum Transducer.

Inside the vessel, a collimated 5.5 MeV Am-241 alpha source with a rate of 200 Hz was suspended by a rotary feedthrough. This allowed us to generate alpha tracks in the active volume by rotating the source in, and to image cosmic muons by rotating it out. The electrons generated in these events were drifted to the top of the TPC by means of a constant electric field, created by an 19 cm-long field cage. Facing this field cage, at the very the bottom of the setup, there was an 8-inch Hamamatsu R5912-02MOD PMT.

At the top of the field cage, two glass-THGEMs with holes of diameter 0.5 mm and pitch 0.8 mm were installed, separated by a transfer gap of 5 mm. The substrate (1 mm thick) of the bottom THGEM was made from Schott Borofloat 33, whilst that of the top THGEM was from fused silica. In both cases, the electrodes were made from an indium tin oxide (ITO) coating on the substrate\cite{glass_thgems}. Once the electrons released by ionizing particles arrive at the THGEMs, they experience a high electric field emanating from the holes. This drags them in, endowing them with enough energy to ionize and excite the gas, a process that generates a considerable amount of photons. Normally, photons produced following argon excitations lie in the so-called $2^{\textnormal{nd}}$ continuum (128 nm) \cite{Santorelli2021,Buzulutskov2018}. The collection of light at this wavelength is usually difficult and requires the use of solid wavelength-shifters such as TPB \cite{Benson2018}, which are subject to ageing and increase the point-spread-function of the system. It has been recently shown through studies of the primary scintillation of Ar/CF$_4$ mixtures at various pressures and concentrations, that CF$_4$ provides strong wavelength shifting into the visible region when at \%-level \cite{Amedo2023}, opening an alternative path.

\subsection{Experimental procedure}

The setup was pumped down to 10$^{-2}$ mbar before filling with a gas mixture of 1\% CF$_4$ and 99\% argon at 1 bar. By carrying out a sequential three-step gas-filling, it was ensured that the pressure gauge was in a species-independent regime and that there was a thorough mixing of the gases.

In order to characterize the optical gain as a function of the voltage applied to the THGEMs, EMCCD data was collected with a long exposure time using an alpha source. When the bottom THGEM was operated, the top one had no electric field applied. In contrast, when the top THGEM was operated, the bottom one had a small electric field to ensure a good electron transmission. In both cases, the background was characterized thoroughly to eliminate any possible light contamination.  The reduced drift field and transfer field between both THGEMs was 0.1 kV/cm/bar and 1 kV/cm/bar respectively, for all measurements.

Once both THGEMs had been characterized, the alpha source was removed in order to collect data from cosmic rays. In this case, both THGEMs were biased simultaneously up to the highest voltage they were able to sustain before discharging. The breakdown voltage was typically higher than the one obtained for alphas, which is to be expected based on the higher ionization density and rate of the alpha source.

\begin{figure}[h!]
\centering % \begin{center}/\end{center} takes some additional vertical space
\includegraphics[width=0.49\textwidth,origin=c,angle=0]{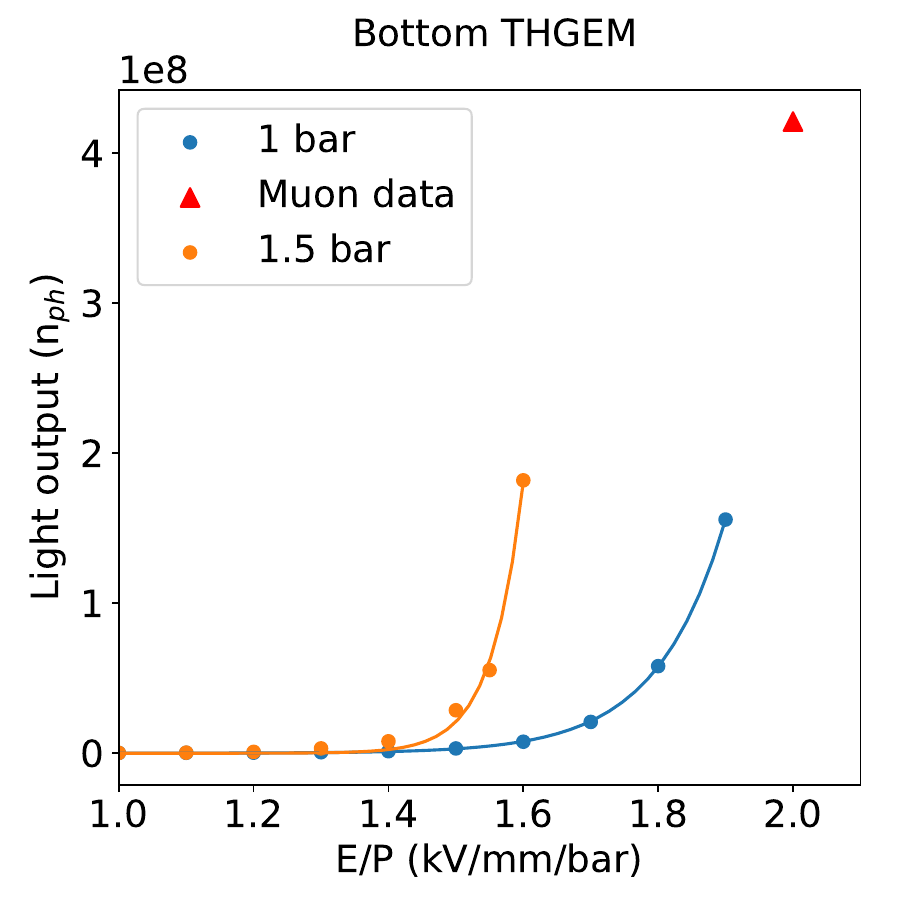}
\includegraphics[width=0.49\textwidth,origin=c,angle=0]{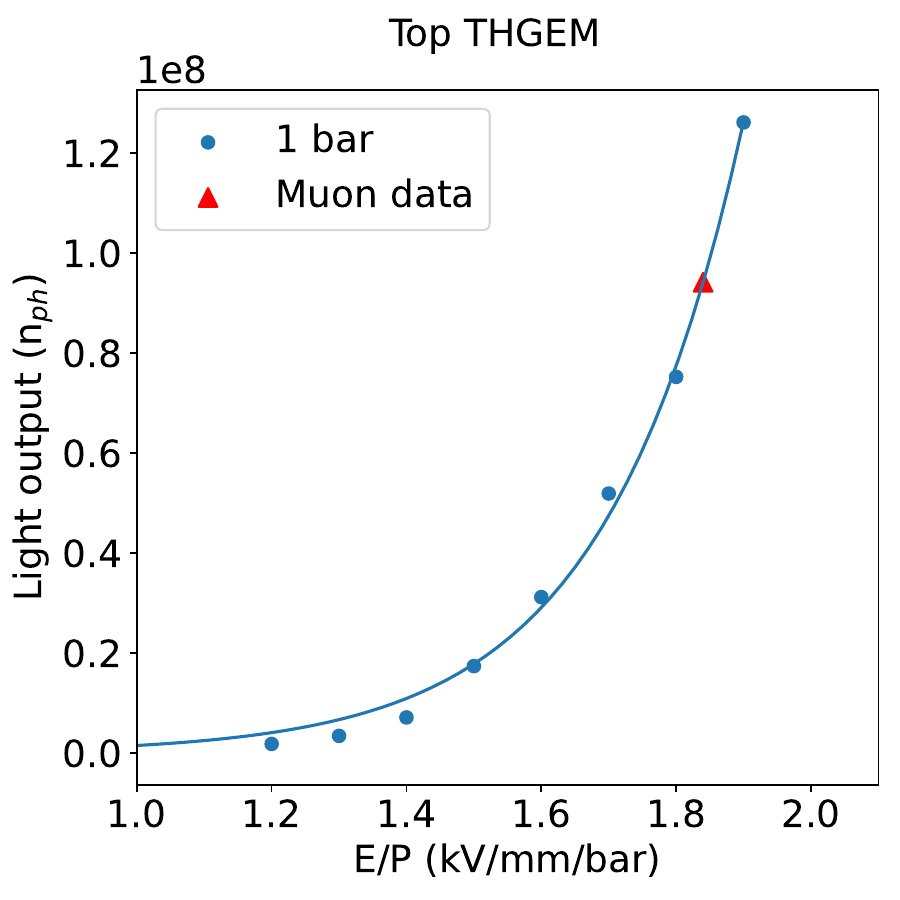}

% "\includegraphics" from the "graphicx" permits to crop (trim+clip)
% and rotate (angle) and image (and much more)
\caption{Light output from the EMCCD camera (n$_{ph}$, number of detected photons) for bottom and top THGEMs at 1 bar (blue) and 1.5 bar (orange) in Ar/CF$_4$ (99/1), as a function of the pressure-reduced electric field across the THGEMs (background light at zero field has been subtracted). The solid lines represent exponential fits to the experimental data. The extrapolation/interpolation (red triangles) of the light output for the conditions at which muon data was taken, that is, 1.84 kV/mm/bar for the top THGEM and 2 kV/mm/bar for the bottom THGEM, is also displayed. No sizeable deterioration of the maximum gain was observed under a modest 50\% increase in pressure. The pressure was limited by the experimental setup.}
\label{Gain_curves}
\end{figure}

\begin{figure}[h!]
\centering % \begin{center}/\end{center} takes some additional vertical space
\includegraphics[width=\textwidth,origin=c,angle=0]{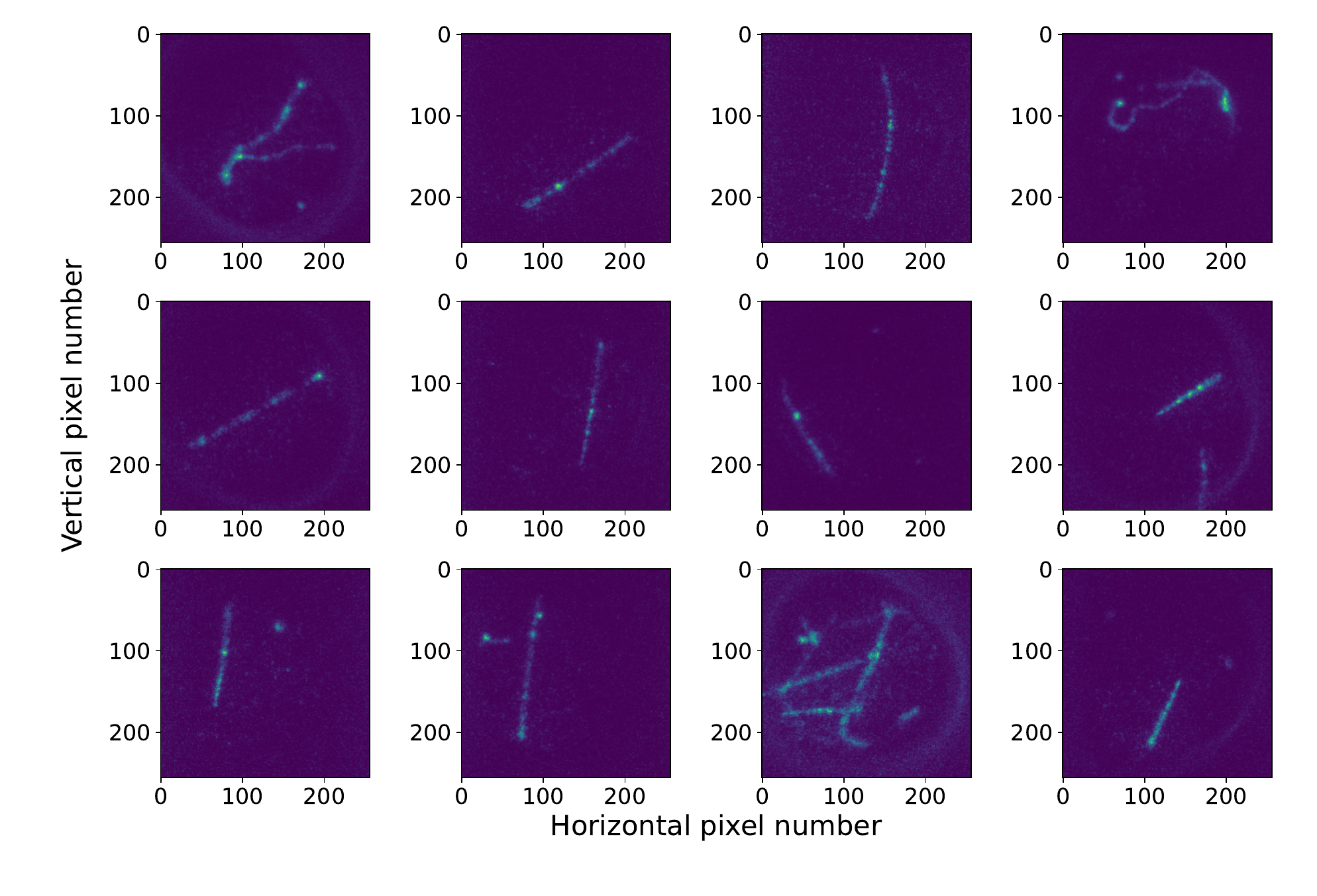}
% "\includegraphics" from the "graphicx" permits to crop (trim+clip)
% and rotate (angle) and image (and much more)
\caption{Gallery of cosmic rays at 1 bar in argon (1\% CF$_4$ doping). The working fields were 1.84 kV/mm/bar for the top THGEM and 2 kV/mm/bar for the bottom THGEM. The exposure time was 0.1s in all cases and the EMCCD gain was set to 1000.}
\label{Cosmics_gallery}
\end{figure}

\section{Analysis and results}
\label{sec:Results}
\subsection{Alpha particles}
\label{subsec:Results_Cosmics}

The integrated light intensity in the alpha-irradiated region of the images from the EMCCD camera can be studied as a function of voltage, for both THGEMs (Fig \ref{Gain_curves}). The optical gain (at the point of production) was estimated from the following formula:

\begin{equation}
gain_{ph} = \left(\frac{n_{ph}W_{i}}{GE\cdot QE\cdot E_{\alpha}\cdot f_{\alpha}\cdot t_{exp} }\right) \!\!\!\!\!\!\
\label{eq:gain}
\end{equation}
by resorting to the following magnitudes: the quantum efficiency (QE) of the EMCCD camera, the geometrical efficiency of the setup (GE), the W$_i$ value of the mixture \cite{Reinking1986}, the exposure time ($t_{exp}$) and the rate ($f_{\alpha}$) and energy ($E_{\alpha}$) of the Am-241 alpha source. In this case, the GE was assumed to be the solid angle subtended by the lens at the object plane and the QE was taken from the manual. When the bottom THGEM was studied, a transparency factor of 75\% was considered. Estimates for bottom and top THGEMs using this method provided a maximum optical gain of the order of $\sim 10^3$ (ph/e) for each of them.

\subsection{Cosmic ray gallery}
\label{subsec:Results_Alphas}

Once the alpha source had been removed from the setup, data from cosmic rays was taken. A selection of different events with varying track topologies can be seen in Fig. \ref{Cosmics_gallery}.

\section{Discussion and conclusions}
\label{sec:Discussion}

Optical readout technologies have improved greatly over the past years, showing good prospects for application to dual-phase liquid argon TPCs \cite{,ARIADNE_beam,ARIADNE_plus}. The addition of the Timepix3 based camera has overcome some of the past caveats of the optical readout, enabling triggerless 3D native readout with ns time resolution and high readout rates \cite{ARIADNE_timepix,ariadne_cf4}. Moreover, this technology greatly reduces readout complexity %and detector cost 
when compared to classical charge readout. It also decouples electrically and mechanically the amplification and readout stages, allowing easy access to the cameras mounted outside the main vessel for maintenance or repairing.

We have demonstrated in this contribution the operation of two stacked glass-THGEMs in an OTPC with an Ar/CF$_4$ mixture (99/1) at around atmospheric pressure. The optical gain of both structures was characterized by using an alpha source and found to be around $\sim 10^3$ per stage, enabling cosmic ray imaging. Although these results have been obtained at around atmospheric pressure (1-1.5~bar), a systematic study as a function of pressure will follow this publication. %These high gains enable the imaging of cosmic ray tracks at low pressure, something far from trivial given the low energy deposition of these particles in a gaseous medium. Considering all the results, we believe the use of this amplification technology will be feasible at higher pressures since the $dE/dx$ will increase linearly with it.

\bibliographystyle{JHEP}
\bibliography{biblio.bib}
\end{document}